\documentclass[12pt]{iopart}
\usepackage{graphicx}
\begin{document}

\title[CSs of P{\"o}schl-Teller
potential and their revival dynamics] {Coherent states of
P{\"o}schl-Teller potential and their revival dynamics}

\author{Utpal Roy, J. Banerji and P.K. Panigrahi
\footnote[3]{   E-mail: prasanta@prl.ernet.in} }

\address{Physical Research Laboratory, Navrangpura,
Ahmedabad, 380 009, INDIA}

\begin{abstract}
     A recently developed algebraic approach for constructing coherent
states for solvable potentials is used to obtain the displacement
operator coherent state of the P\"{o}schl-Teller potential. We
establish the connection between this and the annihilation
operator coherent state and compare their properties. We study the
details of the revival structure arising from different time
scales underlying the quadratic energy spectrum of this system.
\end{abstract}

\section{\large Introduction}
\vskip .5cm
Since its introduction by Schr\"{o}dinger
\cite{schrodinger}, coherent states (CSs) have attracted
considerable attention in the literature
\cite{klauder1,klauder2,klauder3,perelomov,gilmore,GSA}. A variety
of coherent states, e.g., minimum uncertainty coherent state
(MUCS), annihilation operator coherent state (AOCS), displacement
operator coherent state (DOCS) and recently Klauder type CS
\cite{klauder3}, possessing temporal stability, have been
constructed and applied to diverse physical phenomena
\cite{klauder2}. Coherent states of systems possessing nonlinear
energy spectra are of particular interest as their temporal
evolution can lead to revival and fractional revival, leading to
the formation of Schr\"{o}dinger cat and cat-like states
\cite{averbukh,bluhm,robinett}. A celebrated example in quantum
optics of the aforementioned phenomenon is the coherent state in a
Kerr type nonlinear medium \cite{tara}. In quantum mechanical
potential problems, Hamiltonians for potentials like,
P\"{o}schl-Teller, Morse and Rosen-Morse (RM) lead to nonlinear
spectra. Time evolution of the CSs for these potentials, a subject
of considerable current interest
\cite{nietoprd3,benedict,roy,fakhri,shapiro,nieto3,nieto2,crawford,kinani,nieto1,hassouni},
can produce the above type of states.

The simplest way to construct CSs is a symmetry based approach
\cite{perelomov}. It is well-known that making use of the
Heisenberg algebra $\left[a,a^\dag\right]=1$, one can construct
all the above type of CSs for the Harmonic oscillator, which are
identical to each other. In many physical problems, groups like
SU(2) and SU(1,1) manifest naturally, enabling a straightforward
construction of CSs. For the identification of the symmetry
structure of quantum mechanical potential problems, recourse has
been taken to a number of approaches, starting from the
factorization property of the corresponding differential
equations. For the so called shape invariant potentials, super
symmetric (SUSY) quantum mechanics \cite{khare} based raising and
lowering operators have found significant application. A Klauder
type CS, using a matrix realization of the ladder operators, has
also been constructed \cite{klauder3}. The fact that, SUSY ladder
operators act on the Hilbert space of different Hamiltonians, has
led to difficulties \cite{fakhri} in proper operator
identification of the symmetry generators. The ladder operators
have been taken to be functions of quantum numbers, which makes
the corresponding algebraic structure ambiguous. This, in turn,
creates difficulty in establishing a precise connection between
the complete set of states describing the CS and the symmetry of
the potential under consideration. In a number of approaches an
additional angular variable have been employed \cite{alhassid} to
identify SU(1,1) type algebras for describing the infinite number
of states of some of these potentials. Taking advantage of the
shape invariance property, quantum group type algebras have also
been used for describing the Hilbert spaces \cite{aleixo}.

       Recently, a general procedure for constructing CSs for potential
problems have been developed by some of the present authors
\cite{charan}. The approach makes use of novel exponential forms
of the solutions of the differential equations associated with
these potentials for identifying the symmetry generators
\cite{guru1}. No additional variables are introduced and unlike
SUSY based approaches one stays in the Hilbert space of a given
quantum problem while unravelling its symmetry structure. The
present paper makes use of this approach to study the DOCS of the
P\"{o}schl-Teller potential. The primary motivation for
considering the P\"{o}schl-Teller potential is two-fold. First of
all, it has a quadratic spectrum leading to a rich revival
structure for its CS, which can lead to the formation of cat-like
states. Secondly, many other potentials can be obtained from the
P\"{o}schl-Teller potential by appropriate limiting procedure and
point canonical transformations. Hence, the CSs obtained in this
case may have relevance to other potentials. The temporal
evolution, auto-correlation and quantum carpet structures
\cite{averbukh,robinett,loinaz} of the CSs are carefully analyzed
for delineating their structure and various time scales present in
this problem. These properties are then contrasted with the
corresponding ones of the AOCS \cite{charan}.

      The paper is organized as follows. In the following section, we
briefly outline the procedure to identify the symmetry generators
for quantum mechanical potential problems, based on hypergeometric
and confluent hypergeometric equations. These symmetry generators
are then used for constructing the DOCS for general quantum
mechanical potential problems. Dual nature of the DOCS with the
AOCS is algebraically established. In Section 3, the DOCS for the
P\"{o}schl-Teller potential is constructed and its properties
studied. We identify and analyze the various time scales of the
system  in Section 4 and compare the quantum evolution of the CS
with the classical motion. We conclude in Section 5, after
pointing out various directions for further work.

\section{\large Algebraic structure of quantum mechanical potential problems}
\vskip .5cm
As is well-known, the Schr\"{o}dinger equation for a
number of solvable potentials can be connected with the
hypergeometric (HG) and confluent hypergeometric (CHG)
differential equations (DEs). For example, harmonic oscillator,
Coulomb and Morse potentials belong to the CHG class, whereas
P\"{o}schl-Teller and Rosen-Morse belong to the HG class. Below,
we briefly outline the steps of a novel procedure for solving DEs
which connects the solution of a DE with the space of monomials
\cite{guru}. This is subsequently used for identifying the
symmetry generators underlying quantum mechanical potential
problems.

A single variable linear differential equation can be easily cast
in the form,
\begin{equation}
 [F(D)+P(x,d/dx)]y(x)=0
 \label{mainequation}
 \end{equation}
where the first part $F(D)$ is a function of the Euler operator
$D=x d/dx$, possibly including a constant term and $P(x,d/dx)$
contains all other operators present in the DE under study. The
solution can be written in the form,
\begin{equation}
y(x)= C_\lambda \sum_{n=0}^{\infty}(-1)^n
\left[\frac{1}{F(D)}P(x,d/dx)\right]^n x^\lambda\;,
\label{solution}
\end{equation}
with the constraint $F(D)\;x^\lambda=0$ \cite{guru}. Using
Eq.~(\ref{solution}) the polynomial solutions of the HG and CHG
can be written in closed form exponential forms \cite{guru1}:
\begin{equation}
_2F_1(-n,b;c;x)\;=\;(-1)^{n}\,\frac{\Gamma{(b+n)}
\Gamma{(c)}}{\Gamma{(c+n)} \Gamma{(b)}}
e^{\frac{1}{(D+b)}P(x,\frac{d}{dx})} x^n\;, \label{hyper}
\end{equation}
and
\begin{equation}
_1F_1(-n;c;x)=\;(-1)^{n}\,\frac{\Gamma{(c)}}{\Gamma{(c+n)}}
e^{P(x,\frac{d}{dx})} x^n\;. \label{conhyper}
\end{equation}
The exponential forms of these solutions are ideal for identifying
algebraic structures of the solution spaces. For that purpose, one
first identifies raising and lowering operators in the space of
monomials. The operators at the level of polynomials can be
obtained through similarity transformations. The simplest lowering
operators at the level of monomial for CHG and HG functions can be
taken \cite{guru1} as

\begin{equation}\label{shjsjsk}
K_-=x\frac{d^2}{dx^2}+c\frac{d}{dx},\;\;and\;\;
\bar{K}_-=\frac{1}{(D+b)}(x\frac{d^2}{dx^2}+c\frac{d}{dx}),
\end{equation}
respectively.
 The only criterion in choosing these operators at the monomial level is
that, these do not lead to divergent expressions after the
similarity transformation. It can be easily shown that, for the
CHG case, the following generator form a SU(1,1) algebra at the
monomial level:
\begin{equation}\label{CHGalgebra}
K_-=x\frac{d^2}{dx^2}+c\frac{d}{dx},\;\;\; K_+=x,\quad and
\;K_3=x\frac{d}{dx}+\frac{c}{2}.
\end{equation}
Similarly, for the HG case, the SU(1,1) generators are given as,
\begin{eqnarray}
\bar{K}_-&=&\frac{1}{(D+b)}(x\frac{d^2}{dx^2}+c\frac{d}{dx}),\nonumber\\
\bar{K}_+&=&(D+b-1)x,\;\;and\;\;K_0=x\frac{d}{dx}+\frac{c}{2}.
\label{HGalgebra}
\end{eqnarray}

Modulo a normalization, the DOCS for the HG type DE can be
written, at the monomial level, as
\begin{equation}
\Phi^{\beta}(x)=e^{\beta \bar{K}_+} x^0,
\end{equation}
Here, $x^0=1$ is the fiducial state satisfying,
\begin{equation}
\bar{K}_-x^0=0.
\end{equation}

To find the CS $\chi (x,\beta)$ at the level of the polynomial, we
make use of the exponential form of the solution in
Eq.~(\ref{hyper}). The DOCS, $\chi (x,\beta)$, can then be written
as,
\begin{eqnarray}
\chi (x,\beta)\;&=&\; e^{-\bar{K}_{-}} e^{\beta \bar{K}_{+}}
x^0 \nonumber\\
&=&\;e^{-K_{-}}\;\sum_{n=0}^{\infty}\;\frac{\beta^{n}}{n!}\left[(D+b-1)
x \right]^{n}
x^0\nonumber\\
&=&\sum_{n=0}^{\infty}\frac{\beta^{n}}{n!}
\frac{\Gamma{(b+n)}}{\Gamma{(b)}}e^{-K_{-}}x^n\nonumber\\
&=&\sum_{n=0}^{\infty}\frac{\beta^{n}}{n!}
(-1)^n\frac{\Gamma{(c+n)}}{\Gamma{(c)}}\;_2F_1(-n,b,c,x)\;.\nonumber\\
\label{generalcoherentstate}
\end{eqnarray}
It is worth noting that, since the similarity transformation does
not affect the algebraic structure, the SU(1,1) algebras remain
intact at the polynomial level, albeit with different expressions
for the generators.

It is interesting to note that, at the monomial level, the DOCS
found above is nothing but the AOCS of $\tilde{K}_-$, \rm{i.e.}
$\tilde{K}_-\Phi^{\beta}(x)=\beta\Phi^{\beta}(x)$, where
\begin{equation}
\tilde{K}_-\;=\;\frac{1}{(D+b)(D+c)}\;\left(x\frac{d^2}{dx^2}+c\frac{d}{dx}\right).
\end{equation}

One notices that $\left[\tilde{K}_-,\bar{K}_+\right]=1$. Hence,
the above procedure is akin to the oscillator construction of
AOCS. We can also identify a $\tilde{K}_+$, which leads to the
oscillator algebra $\left[\bar{K}_-,\tilde{K}_+\right]=1$:
\begin{equation}
\tilde{K}_+\;=\;\left(\frac{D+b-1}{D+c-1}\right)x.
\end{equation}
The AOCS considered earlier \cite{charan}, is the eigen state of
$\bar{K}_-$ and is of the form $e^{\beta \tilde{K}_+} x^0$. This
relationship between DOCS and AOCS has been referred earlier as
duality of these two type of CSs \cite{shanta}. Thus far, the
specific nature of the potential has not been invoked. Now, we
shall use this form to find out the CS for P\"{o}schl-Teller (PT)
potentials.

\section{\large Coherent state for the
symmetric-P\"{o}schl-Teller potential}

\vskip .5cm The trigonometric P\"{o}schl-Teller potential belongs
to the HG class having an infinite number of bound states. Hence
it is natural to expect an underlying SU(1,1) algebra as its
spectrum generating algebra. In reference \cite{charan} AOCS of
the P\"{o}schl-Teller potential has been constructed, making use
of a novel exponential form of the solution of the hypergeometric
differential equation. Below we will concentrate on the
construction of DOCS, following the same procedure and study its
properties. We also compare the properties of DOCS and AOCS.

The eigen values and eigen functions \cite{quesne} of the
symmetric-P\"{o}schl-Teller potential
\begin{equation}
V_{SPT}(y)\;=\;\frac{\hbar^2\alpha^2}{2m}\left[\frac{\rho(\rho-1)}{\cos^2\alpha
y}\right]\;,\;\;\quad \rho >1, \label{SPT}
\end{equation}
are given by,
\begin{eqnarray}
E_{n}^{SPT}&=&\frac{\hbar^2\alpha^2}{2m} (n+\rho)^2,\quad
n=0,1,2,... ~~~\textrm{and}\nonumber\\
\Psi_{n}^{SPT}(\bar{x})&=&\left[\frac{\alpha(n!)(n+\rho)\Gamma{(\rho)}\Gamma{(2
\rho)}}{\sqrt{\pi}\Gamma{(\rho+\frac{1}{2})}\Gamma{(n+2\rho)}}\right]^\frac{1}{2}
(1-\bar{x}^2)^{\frac{\rho}{2}} C_n^\rho (\bar{x}),\nonumber\\
\label{SPTeigenfunction}
\end{eqnarray}
with $\bar{x}=\sin{\alpha y}$. Using the relation
\cite{gradshteyn}
\begin{equation}
C_{n}^{\rho}(1-2x)\;=\;\frac{\Gamma{(2\rho+n)}}{\Gamma{(2\rho)}\Gamma{(n+1)}}\;
_2F_1(-n,b,c,x)
\end{equation}
and $\bar{x}=1-2x$, where $b=2\rho+n$ and $c=\rho+1/2$, we obtain
from Eq.~(\ref{generalcoherentstate})
\begin{equation}
\chi(\bar{x},\beta)=\sum_{n=0}^{\infty}\;\frac{(-\beta)^{n}}{n!}
\left[\frac{\Gamma{(\rho+n+1/2)}\Gamma{(2\rho)}}{\Gamma{(2\rho+n)}\Gamma{(\rho+1/2)}}\right]
C_{n}^{\rho}(\bar{x}) \label{xchharacoherentstate}
\end{equation}
Now multiplying Eq.~(\ref{xchharacoherentstate}) by
$(1-\bar{x}^2)^\rho/2$ and comparing with
Eq.~(\ref{SPTeigenfunction}), we get the coherent state in energy
eigenfunction basis as,
\begin{equation}
\bar{\chi}(\bar{x},\beta)=\sum_{n=0}^{\infty}\;\;d_{n}
\Psi_n^{\mathrm{SPT}}(\bar{x})\;\;, \label{finalSPT}
\end{equation}
where
\begin{equation}
d_n\;=\;(-\beta)^n
\left[\frac{\Gamma{(\rho+n+1/2)}^2}{\Gamma{(2\rho+n)}\Gamma{(n+1)}(n+\rho)}\right]^{1/2}.
\end{equation}

For comparison, the eigen function distribution for AOCS can be
written as
\begin{equation}
d_n^{AOCS}\;=\;(\gamma)^n
\left[\frac{1}{\Gamma{(2\rho+n)}\Gamma{(n+1)}(n+\rho)}\right]^{1/2}.
\end{equation}
\begin{figure}
\centering
\includegraphics[width=3.5in]{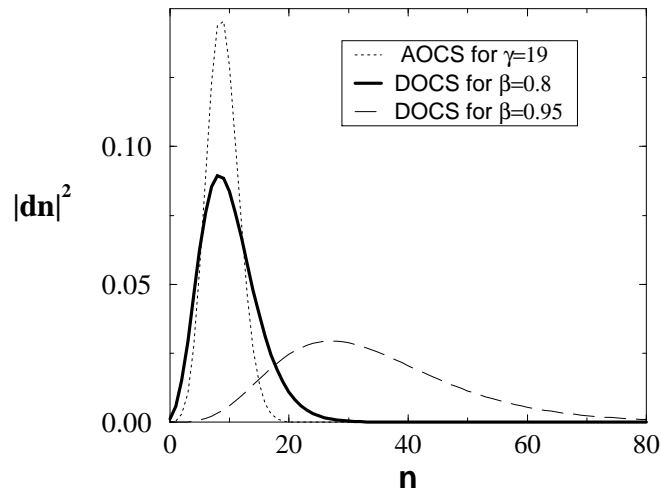}
\caption{The $|d_n|^2$ plots of DOCS and AOCS of
symmetric-P\"{o}schl-Teller potential for
$\rho=15$.}\label{SPTdn2}
\end{figure}
We can also obtain the DOCS of a general trigonometric
P\"{o}schl-Teller potential \cite{klauder3} modulo a normalization
factor, in the same manner as for the symmetric-P\"{o}schl-Teller
case :
\begin{equation}
\chi^{PT}(\bar{x},\beta)\;=\;\sum_{n=0}^{\infty}\;\;d_{n}^{PT}\;\psi_{n}^{PT}(\bar{x})
\end{equation}
where
\begin{eqnarray}
d_{n}^{PT}&=&(-\beta)^{n}\left[\frac{\Gamma{(k+n+1/2)}\Gamma{(\rho+n+1/2)}}{(k+\rho+2
n)\Gamma{(n+1)}\Gamma{(k+\rho+n)}}\right]^{1/2},\nonumber\\
\Psi_{n}^{PT}(\bar{x})&=&\left[\frac{2
\alpha(k+\rho+2n)\Gamma{(n+1)}
\Gamma{(k+\rho+n)}}{\Gamma{(k+n+1/2)}
\Gamma{(\rho+n+1/2)}}\right]^{1/2} \nonumber\\
&\times&(1-\bar{x})^{\rho/2} (\bar{x})^{k/2}
P_{n}^{k-1/2,\rho-1/2} (1-2\bar{x}).
\end{eqnarray}
\vskip.5in
Although the symmetric-P\"{o}schl-Teller potential has
infinite number of bound states, only a few states  contribute
appreciably to the sum, which is peaked around $n=\bar{n}$. In
Fig.~\ref{SPTdn2}, we compare the nature of the distributions of
the eigen states for AOCS and DOCS. For the purpose of comparison,
we have taken the coherence parameters such that, the
distributions are comparable. It is found that, both the eigen
state distributions, peaked at $n=9$, involve the same eigen
states (from n=0 to n=30) for the same potential ($\rho=15$). For
AOCS, the distribution resembles a Gaussian distribution and is
more sharply peaked, as compared to the DOCS. Larger $\beta$ value
makes the distribution flatter for DOCS, as seen in the dashed
curve in Fig.~\ref{SPTdn2}. We now proceed to study the
spatio-temporal dynamics of these wave packets.

\section{\large Revival dynamics of coherent state}
\vskip .5cm
The time evolution of CS $\chi(\bar{x},\beta)$ can be
written as
\begin{equation}
\chi(\bar{x},t)=\sum_{n=0}^{\infty}d_{n}\psi_{n}(\bar{x})e^{-iE_nt}.
\label{timeevolution}
\end{equation}
As the energy expression contains terms up to $n^2$, the system
shows revival and fractional revival but no super-revival
phenomenon. All graphs are plotted in time, scaled by the revival
time $T_{rev}=4\pi/\alpha^2$, .

\begin{figure}
\centering
\includegraphics[width=4in]{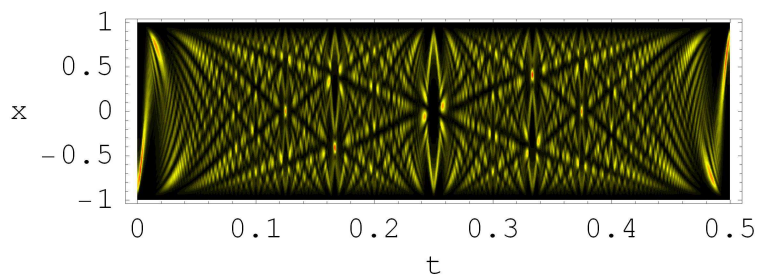}
\caption{(Colour Online) Quantum carpet of the displacement
operator coherent state of symmetric-P\"{o}schl-Teller potential
for $\beta=0.8$ and $\;\rho=10$, brightness signifies the
maximum.}\label{SPTcarpet}
\end{figure}

\begin{figure}
\centering
\includegraphics[width=3.5in]{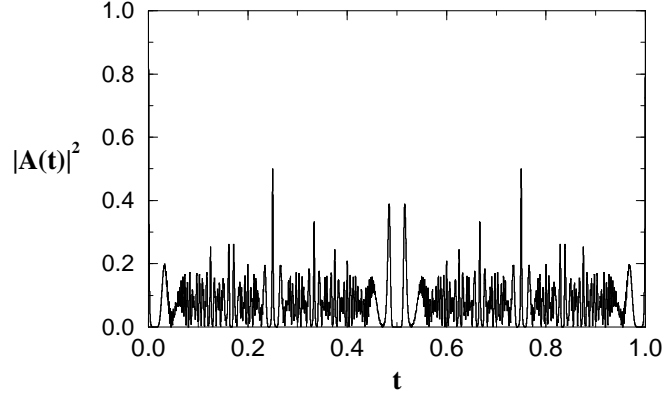}
\caption{Plot of square modulus
$|_{SPT}\langle\chi(x,t)|\chi(x,0.)\rangle_{SPT}|^2$ of the auto
correlation function of DOCS as a function of time, for
$\beta=0.8$, $\rho=10$.}\label{SPTauto}
\end{figure}

In order to throw more light on the structure of the revival
pattern, we note that the eigen functions satisfy
\begin{equation}
\psi_n(-\bar{x})=(-1)^n\psi_n(\bar{x}).
\end{equation}
From Eq.~(\ref{timeevolution}), we can easily obtain the CS wave
packet at time $t=1/2\;T_{rev}$ as
\begin{equation}
|\chi(\bar{x},t=\frac{1}{2}T_{rev})|^2=|\chi(-\bar{x},t=0)|^2.
\end{equation}
\begin{figure*}
\centering
\includegraphics[width=4. in]{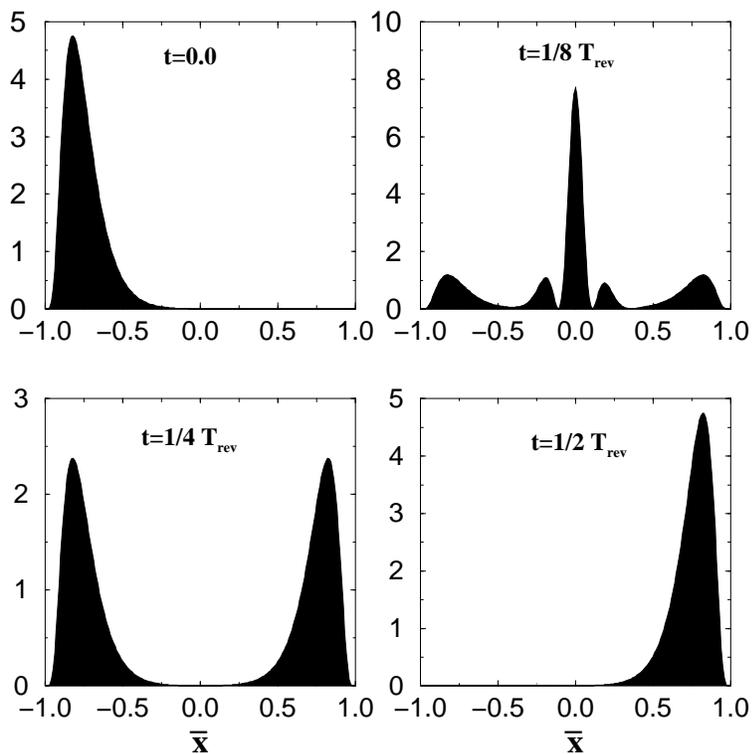}
\caption{Probabity density plot of DOCS of
symmetric-P\"{o}schl-Teller potential at different times for
$\beta=0.8$ and $\rho=10$.} \label{DOCS}
\end{figure*}
\begin{figure*}
\centering
\includegraphics[width=4. in]{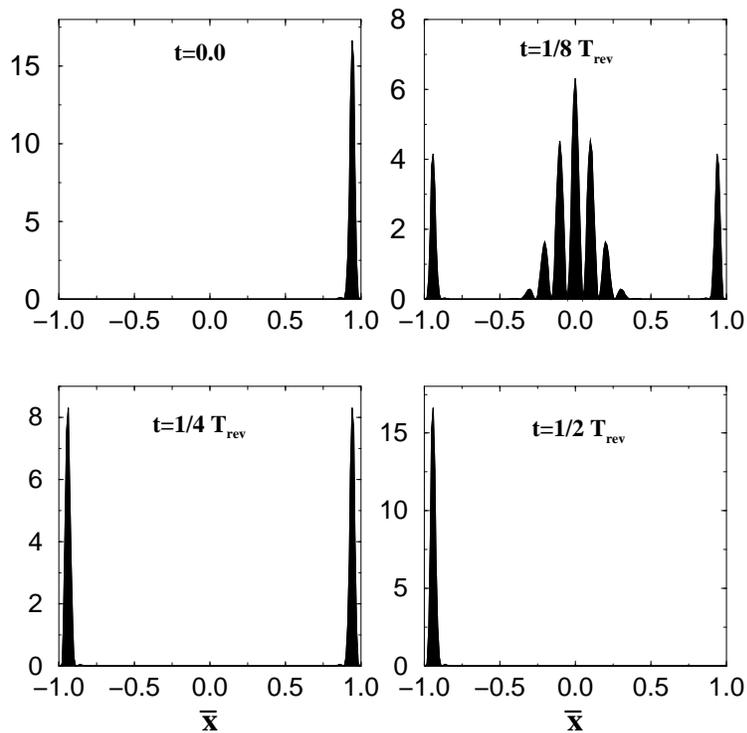}
\caption{Probabity density plot for AOCS of
symmetric-P\"{o}schl-Teller potential at different times for
$\beta^{'}=30$ and $\rho=10$.} \label{AOCS}
\end{figure*}
Thus, at time $t=1/2\; T_{rev}$, a mirror image of the initial
wave packet is produced at the opposite end of the potential well
(Fig.~\ref{DOCS} and \ref{AOCS}). This can be observed as a bright
ray at time $t=1/2\; T_{rev}$, in the quantum carpet structure
(Fig.~\ref{SPTcarpet}). The auto-correlation function
\begin{equation}
A(t)\;=\;\langle\chi(\bar{x},t)|\chi(\bar{x},0)\rangle,
\end{equation}
yields
\begin{equation}
A(t=1/2)\;=\;\sum_{(A+n)\;even}|d_n|^2-\sum_{(A+n)\;odd}|d_n|^2\;.
\end{equation}
As $d_n$ oscillates rapidly, it will not contribute significantly
to the $|A(t)|^2$ (Fig.~\ref{SPTauto}), at $t=1/2\; T_{rev}$. At
time $t=1/4\;T_{rev}$ the CS wave packet becomes
\begin{eqnarray}
\chi(\bar{x},\frac{1}{4} T_{rev})&=&
\frac{1}{\sqrt{2}}e^{-i\pi/4}\left[\chi(\bar{x},0)+e^{i\pi/2}\chi(-\bar{x},0)\right]\nonumber \\
|\chi(\bar{x},\frac{1}{4}T_{rev})|^2\;&=&
\frac{1}{2}\left[|\chi(\bar{x},0)|^2+|\chi(-\bar{x},0)|^2\right]
\end{eqnarray}

In this case, the wave packet breaks up into two parts which are
situated at the two opposite corners of the potential well
(Fig.~\ref{DOCS} and \ref{AOCS}). This gives rise to two bright
spots at the two vertical ends of the quantum carpet at $t=.25$.
At the same instance, the auto-correlation function only gives a
peak, as manifested in Fig.~\ref{SPTauto}. To explain the
probability density plot at $t=0.125$, we consider a fictitious
classical wave packet
\begin{equation}
\chi_{cl}(\bar{x},t)\;=\;\sum_{n=0}^{\infty}\;\;d_{n}\;\psi_{n}(\bar{x})e^{-2\pi
i\frac{n}{T_{cl}}t}\;, \label{classicalwave}
\end{equation}
which behaves like the initial wave packet at small time (order of
$T_{cl}$) where $T_{cl}=\frac{2\pi}{\alpha^2\rho}$. Using the
discrete Fourier transform (DFT), the original CS wave packet
(Eq.~(\ref{timeevolution})) at time $t=\frac{r}{s}T_{rev}$ can be
written as a linear combination of classical wave packet of
Eq.~(\ref{classicalwave}) as
\begin{equation}
\chi(\bar{x},\frac{r}{s}
T_{rev})\;=\;\sum_{p=0}^{l-1}\;\;a_{p}\;\chi_{cl}(\bar{x},\frac{r}{s}
T_{rev}+\frac{p}{l}T_{cl})\;, \label{rswave}
\end{equation}
where
\begin{equation}\label{DFT}
a_p=\frac{1}{l}\;\sum_{n=0}^{l-1}\;\;\exp\left[2\pi
i(n\frac{p}{l}-n^2\frac{r}{s})\right]\;.
\end{equation}

 Here, $r$ and $s$ are two mutually prime integers and $l$
is the period of the quadratic phase term. In general,
$l=\frac{s}{2}$, if $s$ is integral multiple of $4$ and $l=s$ in
all other cases. In this case $\frac{r}{s}=\frac{1}{8}$ and
$l=\frac{s}{2}=4$, Substituting in Eq.~(\ref{rswave}) and
Eq.~(\ref{DFT}), we get
\begin{equation}
\chi(\bar{x},\frac{1}{8} T_{rev})=\;
a_{0}\;\chi_{cl}^{(0)}\;+\;a_{1}
\chi_{cl}^{(1)}\;+\;a_{2}\chi_{cl}^{(2)}\;+\;a_{3}\chi_{cl}^{(3)}\;
\label{chit}
\end{equation}
where
\begin{equation}
\chi_{cl}^{(i)}=\chi_{cl}(\bar{x},\frac{1}{8}
T_{rev}+\frac{i}{4}T_{cl}),\quad i=0,1,2,3\;
\end{equation}
and $a_0=-a_2=\frac{1}{2\sqrt{2}}(1-i)\;$; $a_1=a_3=1/2$.

The above expression (\ref{chit}) signifies that the wave packet
has broken into four parts, each of them differing by a phase
$\pi/4$. In the probability density
\begin{eqnarray}
|\chi(t=\frac{1}{8})|^2\;&=&\;\frac{1}{4}\left[|\chi_{cl}^1|^2\;
+\;|\chi_{cl}^2|^2\;+\;|\chi_{cl}^3|^2\;+\;|\chi_{cl}^4|^2
\right]\\\nonumber
&\;&+\frac{1}{2}\textrm{Re}\left[{\chi_{cl}^1}^*\chi_{cl}^2e^{i\pi/4}-{\chi_{cl}^1}^*\chi_{cl}^3
+{\chi_{cl}^1}^*\chi_{cl}^4e^{i\pi/4}\right.\nonumber\\
&&\;-\left.{\chi_{cl}^2}^*\chi_{cl}^3e^{-i\pi/4}
\;+{\chi_{cl}^2}^*\chi_{cl}^4-{\chi_{cl}^3}^*\chi_{cl}^4e^{i\pi/4}\right],
\end{eqnarray}
the first term carries the contribution from the individual
subsidiary waves and the second term arises due to the
interference between them. We note that the $\chi_{cl}^1$ and
$\chi_{cl}^3$ are spatially well separated, giving very less
contribution in interference. The dominant interference term is
${\chi_{cl}^2}^*\chi_{cl}^4$, as $\chi_{cl}^2$ and $\chi_{cl}^4$
are not spatially separated. Thus, at time $t=0.125$, wave
function splits into four parts, but their interference at the
middle gives a strong peak, rather than giving four distinct
similar waves. For comparison, the wave packet structure for AOCS
of symmetric-P\"{o}schl-Teller potential is also shown in
Fig.~\ref{AOCS}. In this case, as the initial wave packet is
sharper, the interference terms are less dominant than that of
DOCS.

We have observed that the initial wave packet remains in the left
corner of the potential well and oscillates due to the impulse
from the well and at later times, as it spreads, being away from
the boundary of the well. This is quite transparent from the
quantum carpet structure, which gives the space time rays of
probability density of the corresponding coherent states. We note
that, the rays in quantum carpet are not straight lines which is
the case for infinite square-well.

In order to contrast the temporal evolution of the DOCS with the
classical motion, we note that for a particle of energy $E$,
\begin{equation}
x(t)\;=\;a\;\arccos\left[\frac{\alpha_1-\beta_1}{2}+\sqrt{\Delta}\cos{(\sqrt{\frac{2E_c}{m}}\frac{t}{a})}\right]
\end{equation}
where $\Delta=(1-\frac{1}{2}
(\sqrt{\alpha_1}+\sqrt{\beta_1})^{2})(1-\frac{1}{2}(\sqrt{\alpha_1}-\sqrt{\beta_1})^{2})$,
$ V_0=\frac{\alpha^2}{m}$ and
$\alpha_1=\frac{V_0}{E_c}\rho(\rho-1),\;\beta_1=\frac{V_0}{E_c}
k(k-1)$; with the condition
\begin{equation}
E_c>\frac{V_0}{2}(\sqrt{\rho(\rho-1)}+\sqrt{k(k-1)})^2.
\end{equation}
\begin{figure}
\centering
\includegraphics[width=5.5in]{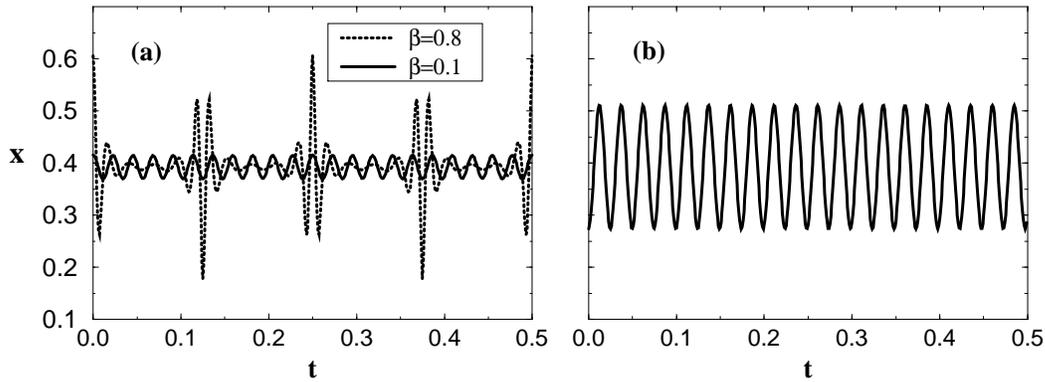}
\caption{Plots of expectation values of x for (a) the coherent
state of the P{\"o}schl-Teller potential, with different values of
$\beta$ where $\rho=5,k=5$ and $\alpha=2$.(b) classical solution
of P{\"o}schl-Teller potential, with $a=0.25,m=1,V_0=4,k=5,\rho=5$
and $E_c=$ average value of energy for $\beta=0.1$.}\label{xpect}
\end{figure}
This classical trajectory is shown in Fig.~\ref{xpect}(b). The
expectation value of the position with respect to the coherent
state of general trigonometric P\"{o}schl-Teller potential is
obtained as
\begin{equation}
\langle x(t) \rangle\;=\frac{1}{\alpha} \arcsin\sqrt{1/2\;(1-z)}
\end{equation}
where
\begin{equation*}
z=N\;\sum_{n=0}^{\infty}\;(2
A_n\;\cos{\left[2\;\alpha^{2}(2n+\rho+k+1)t\right]-C_n})
\end{equation*}
having,
\begin{equation*}
\fl A_n=-(\beta)^{2n+1}\\\left[\frac{2
\Gamma{(\rho+n+3/2)}\Gamma{(k+n+3/2)}}
{\Gamma{(\rho+k+n)}\Gamma{(n+1)}(2n+\rho+k)(2n+\rho+k+1)(2n+\rho+k+2)}\right]
\end{equation*}
and
\begin{equation*}
\fl
C_n=(\beta)^{2n}\left[\frac{
\Gamma{(\rho+n+1/2)}\Gamma{(k+n+1/2)}(k+\rho-1)(k-\rho)}
{\Gamma{(\rho+k+n)}\Gamma{(n+1)}(2n+\rho+k)(2n+\rho+k-1)(2n+\rho+k+1)}\right]
\end{equation*}

N being the normalization constant. This $\langle x(t) \rangle$ is
plotted in Fig.~\ref{xpect}(a) which nearly matches with the
classical trajectory for very small values of $\beta$ (solid line
in Fig.~\ref{xpect}(a)). In this case only a few eigen states
contribute to the coherent state wave packet. Sudden changes in
the $\langle x(t) \rangle$ values are the signatures of revivals
and the fractional revivals \cite{sudheesh}.

\section{\large Conclusions}
\vskip .5cm

 In conclusion, the algebraic procedure used here,
for constructing CSs for potentials based on confluent
hypergeometric and hypergeometric differential equations depends
on the fact that, the solutions of the above differential
equations can be precisely connected with the space of monomials.
This leads to a straightforward identification of symmetry
generators, without taking recourse to additional angular
variables or SUSY type multiple, related Hamiltonians. The nature
of the specific potential enters through the corresponding ground
states and by fixing the parameters and variables of the above
solutions. We have concentrated here on the CS of
P\"{o}schl-Teller potential, since various potentials can be
obtained from the same, through limiting of parameters and point
canonical transformation. The time evolution of the CS for this
potential, having non-linear spectra, produces cat like states in
fractional revivals. We contrasted the properties of the two
different types of CSs possible here, as well as the temporal
evolution of the CS, with classical motion. As has been noted
earlier, this procedure easily extends to more complicated
non-linear coherent states arising from deformed algebras, a
subject we intend to take up in future. We also would like to
analyze the subject of mesoscopic superposition and sub-Planck
scale structure \cite{zurek}, possible in this type of quantum
systems.

\end{document}